\documentclass[runningheads]{llncs}
\usepackage{graphicx,xcolor,cancel}
\usepackage{listings}
\usepackage{color}
\definecolor{cobalt}{RGB}{61,89,171}
\usepackage[colorlinks,citecolor=cobalt,linkcolor=cobalt,urlcolor=cobalt,
pdfpagemode=UseNone,backref = page,linktocpage]{hyperref}

\usepackage{amsmath,amsfonts, amssymb}
\begin{document}
\title{Quantum Register Algebra: 
the mathematical language for quantum computing\thanks{Supported by FAPESP and CNPq.}}
\titlerunning{QRA: 
the
%a 
%new 
%advantageous 
mathematical language for QC}
\author{
Hrdina, J.\inst{1}\orcidID{0000-0003-2460-7571} \and
Hildenbrand D.\inst{2}\orcidID{0000-0002-6384-4345
} \and \\
N\'avrat A.\inst{1}\orcidID{0000-0002-8100-4032}
\and
Steinmetz C.\inst{4}\orcidID{-}
\and \\
Alves R.\inst{3} \orcidID{0000-0002-3171-9339}
\and
Lavor C.\inst{4}\orcidID{0000-0002-8105-3627}
\and \\
Va\v s\'ik P.\inst{1}\orcidID{0000-0001-6725-7378}
\and
Eryganov I.\inst{1}\orcidID{0000-0003-2203-882X}
}
\authorrunning{J. Hrdina et al.}
\institute{Brno University of Technology, Czech Republic \\
\email{\{hrdina, navrat.a,vasik\}@fme.vutbr.cz, xperyga00@vutbr.cz}
\and
Technische Universität Darmstadt, Darmstadt, 64297, Germany\\
\email{dietmar.hildenbrand@gmail.com}
\and
Federal University of ABC, Sao Bernardo, Brazil
\email{alves.rafael@ufabc.edu.br}
\and
:em engineering methods AG, Darmstadt, Germany
\email{christian.steinmetz@e-mail.de}
\and
University of Campinas, Campinas, Brazil
\email{clavor@unicamp.br}}
\maketitle              % typeset the header of the contribution
\begin{abstract}
We introduce Quantum Register Algebra (QRA) as an efficient tool for quantum computing. We show the direct link between QRA and Dirac formalism. We present GAALOP (Geometric Algebra Algorithms Optimizer) implementation of our approach. Using the QRA basis vectors definitions given in Section 4 and the framework based on the de Witt basis presented in Section 5, we are able to fully describe and compute with QRA in GAALOP using the geometric product. We illustrate the intuitiveness of this computation by presenting the QRA form for the well known SWAP operation on a two qubit register.
\end{abstract}
\section{Introduction}

Geometric Algebra proved in recent years that it is a mathematical system covering many other mathematical systems as conventionally used in engineering and physical science, \cite{dl,dorst-07,hil1,per}. Examples are linear algebra, quaternions, Dirac and Pauli matrices, Plucker coordinates to mention only a few. This means, the big advantage of Geometric Algebra is that you do not have to learn different mathematical systems in order to handle various application areas. You simply have to learn only one mathematical system to handle them. Another simplification based on Geometric Algebra is that both objects and operations are handled in one algebra, means you do not have to distinguish, for instance, between points or vectors as objects and matrices for the description of operations on them.

Now, the question is: can Geometric Algebra also advantageously be used for quantum computing? Quantum computing, currently, is based on working with complex numbers, matrices of complex numbers and tensor computing in order to handle operations with arbitrary numbers of qubits.

Some papers on geometric algebras and quantum computing demonstrate application of  geometric algebras $\mathbb G_{4,1}$ (relativistic case) and $\mathbb G_{3,0}$ (nonrealtivistic case), \cite{cm}, geometric algebra $\mathbb G_{n,n}$, \cite{qba1} and finally a complex Clifford algebra $\mathbb C_n$,\cite{hrdina2022quantum}.   

In this paper, we show that quantum computing can simply be based on a Geometric Algebra called QRA (quantum register algebra). In Sect. 2, we show that complex numbers can easily be identified within Geometric Algebra.

The Dirac formalism is a very useful formalism to describe quantum computing. Although it is originally based on complex matrices (see Sect. 3), we show in our definition of QRA according to Sect. 4 and Sect. 5, that it can still be used with QRA.

In order to compute with QRA, we use GAALOP, a stand alone geometric algebra algorithm optimizer [Ref for GAALOP]. We extended the GAALOP tool presented in [1] for the QRA support based on the definitions presented in Sections 4 and 5. The full implementation is shown in Sect. 6. In order to simplify the computing as much as possible, we also integrated it in our online tool GAALOPWeb. In Sect. 7, we present how easy it is to compute with QRA using GAALOP, for this we present the QRA form for the SWAP gate. The big advantage for quantum computing beginners is that they only have to know Geometric Algebra in order to describe the objects and operations of quantum computing.

\section{From complex numbers to geometric algebras and back again}

The algebra of complex numbers $\mathbb C$ is an essential tool for quantum computing. Qubits are realised by vectors in complex vector space $\mathbb C^n$ and the gates by matrices $n \times n$ over the complex numbers. In our approach, the complex linear algebra is the language for quantum computing. More precisely, we are using the natural concept of complex geometric algebras with the complex numbers as one of their instances. In the sequel, we investigate these concepts in a more detailed way.

Formally, geometric algebra $\mathbb G_{p,q}$ is a free, associative, distributive and  unitary algebra over the set of abstract elements $\{e_1, \dots , e_n\}$ endowed with the following identities: 
\begin{align} \label{ident}
%\begin{split} 
e_i e_j &= - e_j e_i \text{, where } i \neq j,\nonumber \\
e_i^2 &=  1 \text{, where } i =1,\dots,p, \\
e_i^2 &= - 1\text{, where } i =p+1,\dots,n.\nonumber
%\end{split}
\end{align}
In computer science notation, we understand  $\mathbb G_{p,q}$ as a vector space where vectors are built as words over the alphabet $\{e_1, \dots , e_n\}$, including an empty word, using the following equivalency. Two words are equivalent (they are different representations of the same object) if the first can be rewritten into the second and vice versa  with the help of identities \eqref{ident}, distributivity and associativity. 

For example, let us consider geometric algebra $\mathbb G_{1,1}$, so we have words over the alphabet $\{ e_1,e_2\} $ and identities $e_1^2=1,\  e_2^2=-1$ together with anti--commutativity $e_1 e_2=-e_2e_1 $. Then the vector space basis is of the form
\begin{align} \label{bas}
\{1, e_1, e_2, e_1e_2 \}.     
\end{align}
Hence,  for example the word $e_2e_1e_2e_1e_1e_2e_2$ can by rewritten to $-e_1$ with the help of identities \eqref{ident} in the following way:
\begin{align*}
e_2e_1e_2e_1e_1e_2e_2 = e_2e_1e_2(1)(-1)
= e_1e_2e_2= e_1(-1)=-e_1.
\end{align*}
The element $e_1$ is in the basis \eqref{bas}.
We can see that the elements of geometric algebra $\mathbb G_{1,1}$  are the linear combinations over the basis \eqref{bas}, i.e.
\begin{align}
\mathbb G_{1,1} = \{x_1+x_2e_1+x_3 e_2+x_4 e_1e_2 | x_i \in \mathbb R \}
\end{align}
together with multiplication given by identities \eqref{ident}.  For example
\begin{align*}
(e_1+2 e_1e_2) ( 1+e_1-3e_2)
&= e_1( 1+e_1-3e_2) +2 e_1e_2( 1+e_1-3e_2) \\
&= e_1+e_1e_1-e_13e_2 + 2 e_1e_2+2 e_1e_2e_1-2 e_1e_23e_2 \\
&= e_1+(1)-3e_1e_2 + 2 e_1e_2- 2  e_2 e_1 e_1 -6 e_1(-1) \\
&= 1+e_1-e_1e_2- 2  e_2 (1) +6 e_1 \\
&= 1+7e_1- 2  e_2 -e_1e_2  \in \mathbb G_{1,1}.
\end{align*}

In any geometric algebra $\mathbb G(p,q)$ such that 
$p>1$ or $q>1$ we can find a subalgebra isomorphically equivalent to $\mathbb C$ in the following way:
if $p>1$ we have two elements $e, f \in \mathbb G(p,q)$
such that $e^2=f^2=1$ and $ef=-fe$, so 
$$(ef)^2=efef=-eeff=-e^2f^2=-1$$
and 
$$\mathbb C \cong \bar{\mathbb C} = \{ a+b (ef) | a,b \in \mathbb R\} \subset \mathbb G(p,q).$$
In the same way, if we have $q>1$ then we have two elements $e,f$ such that $e^2=f^2=-1$ and $ef=-fe$, so 
$$(ef)^2=efef=-eeff=-e^2f^2=-1(-1)(-1)=-1.$$
Note that the element $ef$ commutes with the other basis elements in both cases. 

\begin{remark}
The complex numbers $\mathbb C$ are in fact  isomorphically  equivalent geometric algebra $\mathbb G_{0,1}$. Indeed, $\mathbb G_{0,1}$ is a space  
\begin{align}
\mathbb G_{0,1} = \{x_1+x_2e_1| x_i \in \mathbb R \}
\end{align}
such that $e_1^2=-1$ which is the vector space isomorphically equivalent to  $\mathbb C$.
The specific concept for the imaginary unit used in QRA is described in Sect. 4. 
\end{remark}

\section{Matrices vs. Dirac formalism}
In this section, we show the direct link between Dirac formalism and geometric algebras. Original quantum mechanics is based on Dirac formalism, but quantum computing can be built up of matrices because the Hilbert space of qubits' states is finite. We briefly recall the link between matrices and Dirac formalism, \cite{primer}. 

\begin{itemize}
\item the $n$-qubit (ket)
\begin{align*}
& | i \rangle = | a_1 \cdots  a_{n-1} a_n \rangle, \text{ where } 
a_i \in \{ 0,1\} \text{ and } i=a_1 2^{n-1}+ \cdots +a_n 2^0   \\
&\Longleftrightarrow 
\begin{pmatrix}
0 \\ \vdots \\ 1 \\  \vdots  \\  0
\end{pmatrix},
\text{ where  } 1 \text{ is on {i+1} position, 0 otherwise}. 
\end{align*}
\item the dual $n$-qubit (bra)
\begin{align*}
&\langle i | = \langle a_1 \cdots  a_{n-1} a_n |, \text{ where }
a_i \in \{ 0,1\} \text{ and } i=a_1 2^{n-1}+ \cdots +a_n 2^0  \\
&\Longleftrightarrow 
\begin{pmatrix}
0 & \cdots  & 1 & \cdots 0 
\end{pmatrix},
\text{ where  } 1 \text{ is on {i+1} position, 0 otherwise}. 
\end{align*}
\end{itemize}
The $n$-qubit gates are matrices $A=(a_{ij})$, where $i,j=1,\dots, 2^n$. So if a matrix $A$ acts on coordinates in canonical basis, then its element $a_{ij}$ sends the $j$--th element of canonical basis to $i$--th element of canonical basis in the same way as Dirac expression $|i-1 \rangle \langle j-1|$, so  
$$
\begin{pmatrix}
a_{11} & \dots & a_{1n} \\ \vdots & & \vdots \\
a_{n1} & \dots & a_{nn} 
\end{pmatrix}
\Longleftrightarrow
\sum_{i,j = 1}^n a_{ij}
|i-1 \rangle \langle j-1|.
$$
For example, the representations of 1-qubits (ket and bra) are  
$$
|0 \rangle \Longleftrightarrow \begin{pmatrix} 1 \\ 0
\end{pmatrix},
|1 \rangle \Longleftrightarrow \begin{pmatrix} 0\\ 1
\end{pmatrix},
\langle 0 | \Longleftrightarrow \begin{pmatrix} 1 & 0
\end{pmatrix},
\langle 1 | \Longleftrightarrow \begin{pmatrix} 0& 1
\end{pmatrix}$$
and  the NOT gate is of the form 
$$
\begin{pmatrix}
0 & 1 \\ 1 & 0
\end{pmatrix}
\Longleftrightarrow 
|0 \rangle \langle 1 | + |1 \rangle \langle 0 |.
$$

\section{  Definition of QRA}
Recall that a Geometric Algebra $\mathbb G_{n}=\mathbb G_{n,0}$ is a free, associative, unitary algebra over the set of anti--commuting generators $\{ e_1, \dots , e_n \}$ such that 
$   e_i^2 =1, i \in \{1,\dots,n \} .$ Now we finally define the Quantum Register Algebra (QRA). First, let us consider a geometric algebra $\mathbb G_{n+2}$ with its basis elements $\{ e_1, \dots , e_n, r_1, r_2 \}$ together with the following identities
\begin{align} \label{qra}
e_{1}^2&=e_{2}^2=\cdots =e_{n}^2=r_{1}^2=r_{2}^2=1. 
\end{align}
Then we define a bivector $$\iota = r_1 r_2$$ and show that the set 
$\tilde{\mathbb C} = \{ a+b \iota | a,b \in \mathbb R \} $ is isomorphic to an algebra $\mathbb C$, so $\iota$  plays the role of a complex unit. 
The set $\tilde{\mathbb C}$ is closed with respect to addition and multiplication. The element $\iota$ is in square equal to $-1$. Indeed, 
 $$\iota^2=  r_1 r_2 r_1 r_2 =  -r_1^2 r_2^2 =-1,$$
so $\tilde{\mathbb C}  \cong \mathbb C
$. 
Now we define QRA as a geometric subalgebra $\mathbb G_n$ with the coefficients in $\tilde{\mathbb  C}$, i.e.
$$\mathrm{QRA} = \{ a_0 g_0 + \cdots + a_n g_n  | a_i \in \tilde{\mathbb  C}, g_i \in \mathbb G_n \} \subset \mathbb G_{n+2}.$$
Hence for any element $A \in \mathbb G_{n}$ we have 
$ \iota A = A \iota $ because $\iota$ is a bivector and $\iota \notin \mathbb G_n.$ 

%We will demonstrate our approach on a set of examples....

\section{  QC in the QRA framework}
To use QRA to model quantum computing we choose a different basis of QRA based on geometric algebra $\mathbb G_{2n}$. This basis is called Witt basis and it is formed by elements $\{
f_1, f_1^\dagger, \dots, f_n,f_n^\dagger\}$ satisfying 
\begin{align}
f_i&= \frac{1}{2} (e_i + \iota e_{i+n}) \label{eq:fi}, \\
f_i^\dagger &=\frac{1}{2} (e_i - \iota e_{i+n}) \label{eq:fiT},  \end{align}
where $\iota =r_1r_2$. 
Now, we define an element $I=f_1f_1^{\dagger} \cdots f_nf_n^{\dagger}$ satisfying 
\begin{align}
I^2&=I ,\\
f_i I & = 0, \label{r1}\\
f_i f_i^{\dagger} I & = I \label{r2} 
\end{align}
then 
%We define an element $I=f_1f_1^{\dagger} \cdots f_nf_n^{\dagger}$.
we have a straightforward identification 
of bra and ket vectors with the elements of QRA as follows:
\begin{align} 
\langle a_1 \dots a_n| \hookrightarrow 
I(f_n)^{a_n}
\dots (f_1)^{a_1}, \text{ where }a_i \in \{ 0,1\},\label{ident1}\\
| a_1 \dots a_n \rangle \hookrightarrow 
(f_1^{\dagger})^{a_1} 
\dots (f_n^{\dagger})^{a_n} I, \text{ where }a_i \in \{ 0,1\}.\label{ident2}
\end{align}

\noindent
For example, let us consider the space of 2--qubit states. Then the identification \eqref{ident1} and \eqref{ident2} for the ket vectors reads
\begin{align}
\begin{split}\label{2ket}
|00\rangle & \mapsto  
(f_1^{\dagger})^{0}(f_2^{\dagger})^{0}I= I, \\
|01\rangle & \mapsto  
(f_1^{\dagger})^{0}(f_2^{\dagger})^{1}I= f_2^{\dagger}I, \\
|10\rangle & \mapsto  
(f_1^{\dagger})^{1}(f_2^{\dagger})^{0}I= f_1^{\dagger}I, \\
|11\rangle & \mapsto  
(f_1^{\dagger})^{1}(f_2^{\dagger})^{1}I= f_1^{\dagger}f_2^{\dagger}I.
\end{split}
\end{align} 
So the ket vectors in 2-qubit state space are linear combinations of the basis $\{
f_1, f_1^\dagger, \dots, f_n,f_n^\dagger\}$ elements:
\begin{equation}
|\psi \rangle = 
(\psi_{00} + 
\psi_{01} f_2^{\dagger}  +\psi_{10} f_1^{\dagger}+\psi_{11} f_1^{\dagger}f_2^{\dagger})I.
\label{eq:psi}
\end{equation}

\noindent
To define the quantum gates we have to identify the bra vectors in the similar way: 
\begin{align}
\begin{split}\label{2bra}
\langle 00 | & \mapsto  
I(f_2)^{0}(f_1)^{0}= I, \\
\langle 01| & \mapsto  
I(f_2)^{1}(f_1)^{0}= If_2, \\
 \langle 10| & \mapsto  
I(f_2)^{0}(f_1)^{1}= If_1, \\
\langle 11 | & \mapsto  
I(f_2)^{1}(f_1)^{1}= -If_1f_2.
\end{split}
\end{align} 
Thus the bra vectors in 2-qubit state space are combinations of the basis\\ $\{f_1, f_1^\dagger, \dots, f_n,f_n^\dagger\}$ elements as follows:
$$
\langle \psi | = I(\psi_{00} + 
\psi_{10} f_1  +\psi_{01} f_2-\psi_{11} f_1f_2).
$$

To demonstrate our approach we show the design of the 
SWAP gate \cite{nielsen}. Recall that in Dirac notation,  
%following matrix
%$$\begin{pmatrix}
%1 & 0 & 0& 0\\
%0 & 0 & 1& 0\\
%0 & 1 & 0& 0\\
%0 & 0 & 0& 1
%\end{pmatrix},$$
the SWAP gate is represented  
by 
$$|00 \rangle  \langle 00|+|01  \rangle  \langle 10|+ | 10 \rangle  \langle 01|+|11  \rangle  \langle 11|.
$$
Using the identification \eqref{2ket} and \eqref{2bra}, the SWAP gate may be rewritten as:
\begin{align}
\text{SWAP}&=|00 \rangle  \langle 00|+|01  \rangle  \langle 10|+ | 10 \rangle  \langle 01|+|11  \rangle  \langle 11| \notag \\
&=f_1 f_1^{\dagger}f_2 f_2^{\dagger}+ f_1^{\dagger} f_2 - f_1 f_2^{\dagger}
+ f_1^{\dagger} f_1 f_2^{\dagger} f_2.
\label{eq:swap}
\end{align} 
%because of $II^{\dagger}= II=I$ and $I$ commute with any element.  

Before we present how the gate acts on 2-qubits, let us mention the rules for calculations with the Witt basis $\{ f_i , f_i^{\dagger}\}$ elements in the form of a list of properties which can be verified by straightforward computations:
\begin{align}
(f_i)^2&=(f_i^{\dagger})^2=0 \label{p1} \\
f_i f_j &= -f_j f_i \label{p2},  \quad
f_i^{\dagger} f_j^{\dagger} = -f_j^{\dagger} f_i^{\dagger}   \\
f_i f_i^{\dagger} f_i&=f_i,  \label{p4} \qquad
f_i^{\dagger} f_i f_i^{\dagger}=f_i^{\dagger}
\end{align}
We present the SWAP gate functionality on 2-qubits step by step. 
%Note that the gate works correctly if the element $I$ is attached, i.e. we have to consider  
%\begin{equation}
%\text{SWAP} |\boldsymbol{\psi}\rangle I.
%\label{eq:psiI}
%\end{equation}
%Consider expression $ r_1 \dots r_p, r_i \in \{ f_, f_i^{\dagger} \}$
%\begin{align}
%\text{ if } (r_i=f_j) \text{ and } (f_j^{\dagger} \neq r_k), k>i \text{ then } r_1 \dots r_p \text{ vanishes } \label{r1} \\
%\text{ if } (r_i=f_j) \text{ and } (r_{i+1}=f_j^{\dagger}) \text{ then } f_jf_j^{\dagger} =1
%\label{r2}
%\end{align}
\noindent
Thus we can calculate:
\begin{align*}
&\text{SWAP} |\boldsymbol{\psi}\rangle  =\\
&=(
f_1 f_1^{\dagger}f_2 f_2^{\dagger}
+ f_1^{\dagger} f_2 
- f_1 f_2^{\dagger}
+ f_1^{\dagger} f_1 f_2^{\dagger} f_2
) (\psi_{00} + 
\psi_{01} f_2^{\dagger}  +\psi_{10} f_1^{\dagger}+\psi_{11} f_1^{\dagger}f_2^{\dagger})I
\\
&=f_1 f_1^{\dagger}f_2 f_2^{\dagger} (\psi_{00} + 
\cancel{\psi_{01} f_2^{\dagger}}  +\cancel{ \psi_{10} f_1^{\dagger}}+\cancel{ \psi_{11} f_1^{\dagger}f_2^{\dagger}})I \text{ by \eqref{p1}} \\
&\phantom{=}+ f_1^{\dagger} f_2 (\cancel{\psi_{00}} + 
\psi_{01} f_2^{\dagger}  + \cancel{ \psi_{10} f_1^{\dagger}}+\cancel{ \psi_{11}  f_1^{\dagger}f_2^{\dagger}})I
\text{ by \eqref{r1} and  \eqref{p1}} \\
&\phantom{=}- f_1 f_2^{\dagger}(\cancel{\psi_{00}} + 
\cancel{\psi_{01} f_2^{\dagger}}  +\psi_{10} f_1^{\dagger}+\cancel{ \psi_{11}  f_1^{\dagger}f_2^{\dagger}})I
\text{ by \eqref{r1} and  \eqref{p1}}
\\
&\phantom{=}+ f_1^{\dagger} f_1 f_2^{\dagger} f_2(\cancel{\psi_{00}} + 
\cancel{ \psi_{01} f_2^{\dagger}}  + \cancel{ \psi_{10} f_1^{\dagger}}+\psi_{11} f_1^{\dagger}f_2^{\dagger})I
\text{ by \eqref{r1} and  \eqref{p1}}
\\
&=\underline{(f_1 f_1^{\dagger}f_2 f_2^{\dagger}
)} (\psi_{00} ) I
+(
 f_1^{\dagger} \underline{f_2} ) ( 
\psi_{01} \underline{f_2^{\dagger}} ) I
\text{ by \eqref{r2}}
\\
&\phantom{=}
 +  f_2^{\dagger} \underline{f_1} \psi_{10} \underline{f_1}^{\dagger} I
 +( f_1^{\dagger} f_1 f_2^{\dagger} f_2
 ) (\psi_{11} f_1^{\dagger}f_2^{\dagger})I 
 \\
&=(\psi_{00}  +\psi_{01} f_1^{\dagger} 
 +  \psi_{10} f_2^{\dagger} +
\psi_{11} ( f_1^{\dagger} f_1 f_1^{\dagger} )(f_2^{\dagger} f_2  f_2^{\dagger}) )I
  \text{ the last step by \eqref{r2}}\\
&=(\psi_{00}  +\psi_{01} f_1^{\dagger} 
 +  \psi_{10} f_2^{\dagger} +
\psi_{11} f_1^{\dagger}f_2^{\dagger})I,
\end{align*}
where $\cancel{g}$ and $\underline{g}$ stand for $g=0$ and $g=1$.
%using \eqref{p1} and  \eqref{r1}, respectively.
Finally,  by means of \eqref{2ket}, the final 2-qubit can be rewritten in Dirac notation as 
$$
(\psi_{00} | 00 \rangle +\psi_{01} | 10 \rangle   +  \psi_{10} | 01 \rangle +
\psi_{11} | 11 \rangle)I
$$
which is the expected result.

The other gates may be interpreted in the very same way and thus we showed how quantum computation in QRA is realized. We used the Witt basis axioms \eqref{p1}--\eqref{p4} together with additional  axioms \eqref{r1} and \eqref{r2}. The use of the additional axioms may seem redundant and complicated but they only simplify the written form of calculations and thus the functionality is easier for demonstration. Indeed, these rules may be avoided completely for which we point out the following two reasons:
\begin{itemize}
\item If we interpret the element $I$ as  $f_1f_1^{\dagger} \cdots f_nf_n^{\dagger}$, the rules \eqref{r1} and \eqref{r2} do not apply because they are simple consequences of \eqref{p1}--\eqref{p4}. Only the expressions will be longer.
\item We are using the axioms \eqref{p1}--\eqref{p4} for calculations in the Witt basis which naturally corresponds to Dirac formalism. But Witt basis is just a different set of generators for QRA elements. Thus the axioms \eqref{p1}--\eqref{p4} are derived from geometric algebra axioms \eqref{qra} which are very simple. 
\end{itemize}

Our approach is based on the fact that QRA and the Witt basis provide nice language for written QC schemes. The expressions are very similar to Dirac formalism, so they are very simple to understand for people not familiar with geometric algebra. But unlike abstract Dirac formalism, our objects are elements of QRA which is a geometric algebra based on very simple axioms and, furthermore, it is very easy for implementation as shown in the following section.

\section{ GAALOP implementation}
The integration of QRA into GAALOP is done based on the file Definition.csv according to Chapt. 9 of 
\cite{hil3}.  Listing \ref{lstDefinitionQRA} shows the file for the definition of one qubit.

\begin{lstlisting}[caption={Definition.csv for QRA for one qubit},
label=lstDefinitionQRA]
1,e1,e2,er1,er2

1,e1,e2,er1,er2
e1=1,e2=1,er1=1,er2=1
                                  
\end{lstlisting}

In general, this file consists of 5 lines for the definition of the algebra. In the case of QRA, the lines 2 and 5 are left blank since  the used basis in Line 1 and the standard basis in Line 3 are the same and no transformations between the two bases are needed. The basis is defined by the basis vectors e1, e2, er1 and er2 according to $e_1, e_2, r_1, r_2$ as defined in the previous sections. All their squares are defined to 1 according to line 4.

For each additional qubit we need two additional basis vectors, while er1 and er2 are always the same. The definition of a register with two qubits is shown in Listing \ref{lstDefinitionQRA2}. 

\begin{lstlisting}[caption={Definition.csv for QRA based on two qubits},
label=lstDefinitionQRA2]
1,e1,e2,e3,e4,er1,er2

1,e1,e2,e3,e4,er1,er2
e1=1,e2=1,e3=1,e4=1,er1=1,er2=1
                                
\end{lstlisting}

For the second qubit we need the additional basis vectors e3 and e4. For this paper, we also made a new version of GAALOPWeb\footnote{http://www.gaalop.de/gaalopweb/}. It allows online computations with a number of n qubits without the installation of a specific software.
\section{Example in GAALOP}
As an example we implemented the SWAP gate on GAALOP for a register with two qubits. Following the Equations (\ref{eq:fi}) and (\ref{eq:fiT}), the vectors $f_i$ and $f_i^{+}$ as
\begin{eqnarray*}
&f_1 = \frac{1}{2}(e_1+ \iota e_3),\quad
f_1^{\dagger} = \frac{1}{2}(e_1- \iota e_3)& \\
&f_2 = \frac{1}{2}(e_2+\iota e_4),\quad
f_2^{\dagger} = \frac{1}{2}(e_2+\iota e_4)&,
\end{eqnarray*} 
where $\iota = r_1r_2$. Then, we use the definitions in \eqref{2ket} to implement the basis elements, Equation (\ref{eq:psi}) for $| \psi \rangle$ and (\ref{eq:swap}) to define the operator of the SWAP gate. 
%Following the expression in (\ref{eq:psiI}), 
We apply the SWAP operator on the element $|\psi\rangle$.
%, where $I =f_1^{\vphantom{-1}}f_1{+}f_2^{\vphantom{-1}}f_2^{\dagger}$ satisfies (\ref{r1}) and (\ref{r2}). 
But first, let us see the coordinates of each ket vector basis when they are multiplied by $I$ separately:
\begin{lstlisting}[caption={SWAP gate in QRA for two qubits.},
label=lstCode1]
// Imaginary unit
i = er1*er2;
// Witt basis
f1 = 0.5*(e1+i*e3);
f1T = 0.5*(e1-i*e3);
f2 = 0.5*(e2+i*e4);
f2T = 0.5*(e2-i*e4);
// Element "I"
Id = f1*f1T*f2*f2T;
// ket basis vectors multiplied by "Id"
?ket00 = 1*Id;
?ket01 = f2T*Id;
?ket10 = f1T*Id;
?ket11 = f1T*f2T*Id;
\end{lstlisting}
This is important to show which coordinates correspond to each vector in order to see the amplitudes interchanging when SWAP is applied to some linear combination of these vectors. The output for this code is shown bellow:
\begin{lstlisting}[caption={Basic elements multiplied by $I$.},
label=lstOutputCode1]
ket00[0] = 0.25; // 1.0
ket00[42] = 0.25; // e1 ^ (e2 ^ (e3 ^ e4))
ket00[50] = -0.25; // e1 ^ (e3 ^ (er1 ^ er2))
ket00[55] = -0.25; // e2 ^ (e4 ^ (er1 ^ er2))
ket01[2] = 0.25; // e2
ket01[26] = -0.25; // e1 ^ (e3 ^ e4)
ket01[41] = -0.25; // e4 ^ (er1 ^ er2)
ket01[59] = 0.25; // e1 ^ (e2 ^ (e3 ^ (er1 ^ er2)))
ket10[1] = 0.25; // e1
ket10[32] = 0.25; // e2 ^ (e3 ^ e4)
ket10[40] = -0.25; // e3 ^ (er1 ^ er2)
ket10[60] = -0.25; // e1 ^ (e2 ^ (e4 ^ (er1 ^ er2)))
ket11[7] = 0.25; // e1 ^ e2
ket11[16] = -0.25; // e3 ^ e4
ket11[51] = -0.25; // e1 ^ (e4 ^ (er1 ^ er2))
ket11[54] = 0.25; // e2 ^ (e3 ^ (er1 ^ er2))
\end{lstlisting}
We remark that all identities given in (\ref{p1})-(\ref{r2}) can be observed in GAALOP.
Now, let us choose a vector $|\psi\rangle$ \footnote{The vector $|\psi\rangle$ is not normalized, so the final result should be divided by $\sqrt{29}$.}given by:
\begin{eqnarray*}
|\psi\rangle &=& (|00\rangle+2|01\rangle+3|10\rangle+4|11\rangle)I\\
&=& (1+3f_2^{\dagger} + 3f_1^{\dagger} +4f_1^{\dagger}f_2^{\dagger})I,
\end{eqnarray*}
and apply the SWAP gate on it. The expected result is given by:
\begin{eqnarray*}
|\bar{\psi}\rangle &=& (1+3f_2^{\dagger} + 2f_1^{\dagger} +4f_1^{\dagger}f_2^{\dagger})I\\
 &=&(|00\rangle+3|01\rangle+2|10\rangle+4|11\rangle)I.
\end{eqnarray*}
We add 3 lines to the code in \ref{lstCode1}:
\begin{lstlisting}[caption={Definition and application of the SWAP gate on GAALOP.},
label=lstCode2]
//SWAP
SWAP=(f1*f1T*f2*f2T)+(f1T*f2)-(f1*f2T)+(f1T*f1*f2T*f2);
?psi = ket00 + 2*ket01 + 3*ket10 + 4*ket11;
?SwapPsi = SWAP*psi;
\end{lstlisting}
The output is shown below and then we analyze the result.
\begin{lstlisting}[caption={Output of the application of the SWAP gate on GAALOP.},
label=lstOutputCode2]
psi[0] = 0.25; // 1.0
psi[1] = 0.75; // e1
psi[2] = 0.5; // e2
psi[7] = 1.0; // e1 ^ e2
psi[16] = -1.0; // e3 ^ e4
psi[26] = -0.5; // e1 ^ (e3 ^ e4)
psi[32] = 0.75; // e2 ^ (e3 ^ e4)
psi[40] = -0.75; // e3 ^ (er1 ^ er2)
psi[41] = -0.5; // e4 ^ (er1 ^ er2)
psi[42] = 0.25; // e1 ^ (e2 ^ (e3 ^ e4))
psi[50] = -0.25; // e1 ^ (e3 ^ (er1 ^ er2))
psi[51] = -1.0; // e1 ^ (e4 ^ (er1 ^ er2))
psi[54] = 1.0; // e2 ^ (e3 ^ (er1 ^ er2))
psi[55] = -0.25; // e2 ^ (e4 ^ (er1 ^ er2))
psi[59] = 0.5; // e1 ^ (e2 ^ (e3 ^ (er1 ^ er2)))
psi[60] = -0.75; // e1 ^ (e2 ^ (e4 ^ (er1 ^ er2)))
SwapPsi[0] = 0.25; // 1.0
SwapPsi[1] = 0.5; // e1
SwapPsi[2] = 0.75; // e2
SwapPsi[7] = 1.0; // e1 ^ e2
SwapPsi[16] = -1.0; // e3 ^ e4
SwapPsi[26] = -0.75; // e1 ^ (e3 ^ e4)
SwapPsi[32] = 0.5; // e2 ^ (e3 ^ e4)
SwapPsi[40] = -0.5; // e3 ^ (er1 ^ er2)
SwapPsi[41] = -0.75; // e4 ^ (er1 ^ er2)
SwapPsi[42] = 0.25; // e1 ^ (e2 ^ (e3 ^ e4))
SwapPsi[50] = -0.25; // e1 ^ (e3 ^ (er1 ^ er2))
SwapPsi[51] = -1.0; // e1 ^ (e4 ^ (er1 ^ er2))
SwapPsi[54] = 1.0; // e2 ^ (e3 ^ (er1 ^ er2))
SwapPsi[55] = -0.25; // e2 ^ (e4 ^ (er1 ^ er2))
SwapPsi[59] = 0.75; // e1 ^ (e2 ^ (e3 ^ (er1 ^ er2)))
SwapPsi[60] = -0.5; // e1 ^ (e2 ^ (e4 ^ (er1 ^ er2)))
	\end{lstlisting}
Note that the groups of coordinates $\{0,42,50,55\}$ and $\{7,16,51,54\}$, which refer to the coefficients of $|00\rangle$ and $|11\rangle$ respectively, remained unchanged from \verb|psi| to \verb|SwapPsi|, as expected. However, note that the coefficients in the coordinates $\{2,26,41,59\}$ of \verb|psi| that corresponds to $|01\rangle$ was exactly 2 times the coefficients of \verb|ket01| given by \ref{lstOutputCode1} and, in \verb|SwapPsi|, they became 3 times those same coefficients. On the other hand, the coordinates $\{1, 32, 40, 60\}$ changed from 3 (in \verb|psi|) to 2 (in \verb|SwapPsi|) times the coefficients of \verb|ket10| in \ref{lstOutputCode1}, that corresponds to $|10\rangle.$ So, the output vector \verb|SwapPsi| is exactly the vector $|\bar{\psi}\rangle$.

%\section{ Benefits of QRA}
\section{Conclusion}

In this paper, we presented how natural Geometric Algebra can be used for quantum computing. Based on the newly developed QRA and its support by the extended GAALOP tool, the handling of quantum computing is strongly simplified, especially for beginners. The big advantage for quantum computing beginners is that they only have to know Geometric Algebra in order to describe the objects and operations of quantum computing. We hope that this can be the reason for Geometric Algebra to become "the" language for quantum computing in the future.





% LNCS Homepage, \url{http://www.springer.com/lncs}. Last accessed 4
% \end{thebibliography}
\end{document}